\documentclass{article}
\usepackage{fortschritte}
\usepackage{epsf}
\usepackage{amsmath}
\usepackage{amssymb}
\usepackage{bbm}
\usepackage{rotate}
\newcommand{\pV}{\mathbbm{V}}

\newcommand{\rmd}{\mbox{\rm{d}}}
\newcommand{\e}{\mbox{\rm{e}}}

\newcommand{\qd}{\dot{q}}
\newcommand{\pd}{\dot{\phi}}
\newcommand{\ad}{\dot{\alpha}}

\newcommand{\ac}{Q}

\def\be{\begin{equation}}                     %
\def\ee{\end{equation}}                       %
\def\bea{\begin{eqnarray}}                     
\def\eea{\end{eqnarray}}                       
\begin {document}
\begin{center} 
\hfill FSU-TPI-01/05 \\ 
\hfill ITP-UU-05/02 \\
\hfill SPIN-05/01 
\end{center}           
\def\titleline{
%
%
%
Conifold Cosmologies in IIA String Theory\footnote{Work supported by the `Schwerpunktprogramm Stringtheorie' of the DFG.}
\footnote{Talk given by T.M.\ at the 37th International Symposium Ahrenshoop, Berlin, August 23-27, 2004.}
}
\def\email_speaker{
{\tt 
%
%
Thomas.Mohaupt@uni-jena.de             
}}
\def\authors{
%
%
%
%
%
Thomas Mohaupt\1ad and Frank Saueressig\2ad                           
}
\def\addresses{
%
%
%
%
\1ad       
Institute of Theoretical Physics, Friedrich-Schiller-University, Jena, \\ 
 Max-Wien-Platz 1, D-07743 Jena, Germany \\          
\2ad  
Institute for Theoretical Physics \& Spinoza Institute, \\
Utrecht University, Postbus 80.195, 3508 TD Utrecht, The Netherlands \\        
}
\def\abstracttext{
%
%
We discuss the extension of our recent work on M-theory conifold
cosmologies to general conifold transitions and type-IIA string theory.
}
\large
\makefront
\section{Introduction}
\setcounter{footnote}{0}

Recently a lot of effort has been made to understand the dynamics of string and M-theory compactifications in regions of the moduli space where extra light states occur. If string or M-theory is compactified on a special holonomy
manifold, one usually obtains a 
moduli space of vacua, corresponding to the deformations of the internal 
manifold $X$. For theories with eight or less supercharges this moduli space includes special points where submanifolds of $X$ are contracted to zero 
volume so that $X$ becomes singular.  The full string or M-theory, however,
is still non-singular, due to the winding modes of strings
and branes around the vanishing cycles. Since these modes become 
massless when $X$ becomes singular, we will refer to the corresponding
points in moduli space as ESP (=
extra species point(s), following \cite{beauty}). If the singularities
of $X$ can be resolved in more than one way, this gives rise to
topological phase transitions which connect two families of smooth CY$_3$, 
$X$ and $\tilde{X}$, 
with different topologies.\footnote{We refer to 
\cite{GreRev} for a review and more references.} 

Singularities of $X$ lead to singularities or discontinuities of
the low energy effective action (LEEA) which includes the generically massless
modes of the dimensionally reduced string or M-theory, only. It has been
shown in \cite{Str1,Witten1} that for type-II and M-theory
compactifications on Calabi-Yau threefolds (CY$_3$) these singularities
can be interpreted as arising from illegitimately integrating out
 massless string or brane winding modes. In order to have 
a well defined LEEA, and also to study the dynamics of all the relevant, 
i.e., light modes in the vicinity of the ESP, it is desirable to 
work with an extended LEEA, which includes the additional light modes
as dynamical degrees of freedom. 
The search for such LEEA was initiated in 
\cite{SU2,MohProc}, who considered the case of $SU(2)$-enhancement in M-Theory compactified on a CY$_3$. Subsequent work has covered $SU(2)$-enhancement in heterotic string compactifications on $K3 \times T^2$
\cite{LMZ}, toroidal string compactifications close to the selfdual radius 
\cite{SW}, and also flop \cite{flop,FS3} and 
conifold transitions \cite{CFL,DIS,Lukas:2004du}.

In the latter cases investigating the dynamics of cosmological solutions close to the ESP
has lead to a uniform picture, independently of whether one uses a
non-supersymmetric approximation \cite{flop,SW,Lukas:2004du} or a full 
supergravity LEEA \cite{FS4,Mohaupt:2004pr,DIS}. 
The extended LEEA include a scalar potential which encodes the masses of the
additional light modes. This potential generically leads to a dynamical 
stabilization of the moduli in the transition region. In \cite{Mohaupt:2004pr} this trapping was related to an interplay between Hubble friction occurring in an expanding universe and the shape of the scalar potential. A further trapping mechanism, based on the
quantum production of light particles close to an ESP was discussed in
\cite{beauty}. This indicates that the moduli trapping close to an ESP is a generic
property of string compactifications.

In the following we will focus on the construction of extended LEEA
describing flop and conifold transitions in M-theory and IIA string 
compactifications on a CY$_3$ $X$. In this case the generically massless spectrum 
consists of the ${\cal N}=2$ supergravity multiplet (8  supercharges), 
together with abelian vector and neutral hypermultiplets. The 
additional light modes arising in the vicinity of the ESP are charged hypermultiplets, whose masses
are controlled by vector multiplet scalars. While in flop transitions
the vacuum manifold of the scalar potential only has a Coulomb branch,
the potentials describing conifold transitions have a second vacuum branch,
along which part of the abelian gauge group is Higgsed. 

The basic strategy to include these additional hypermultiplets in the extended 
LEEA is to start with the most general `macroscopic' (lower-dimensional)
gauged supergravity action and to impose constraints derived from the underlying `microscopic' 
(ten- or eleven-dimensional) string or M-theory. While the numbers
of the generically massless neutral vector and hypermultiplets along
the Coulomb (Higgs) branch are determined by the Hodge numbers
of $X$ ($\tilde{X}$), the number of light charged hypermultiplets $N$ equals the 
number of holomorphic curves ${\cal C}_i$
of $X$ which are contracted
to zero volume at the transition point. The reason is that the 
charged hypermultiplets correspond to the winding modes of D2-branes
(type-IIA theory) or M2-branes (M-theory) around the 
${\cal C}_i$.\footnote{See \cite{Str1,Lukas:2004du} for a discussion of the
type-IIB description.} As we will discuss in more detail below,
the charges of the hypermultiplets are determined by the homology
classes of the ${\cal C}_i$. In principle the full string or M-theory
also determines all the (field-dependent) couplings in the LEEA, but 
their computation from the microscopic theory
has not yet been achieved. The only couplings which are known
exactly are the vector multiplet couplings of the five-dimensional (5d) LEEA 
arising in M-theory compactifications on $X$ \cite{SU2,CFL}, where the results
of \cite{Witten1} can be used. The vector multiplet couplings
of the 4d LEEA of type-IIA theory on $X$ are more
complicated. Here integrating out charged hypermultiplets generates
logarithmic singularities in the couplings \cite{Str1} rather than just
discontinuities. However, for the quite similar case of $SU(2)$-enhancement
in heterotic theories on $K3 \times T^2$ the vector multiplet couplings
have been computed in \cite{LMZ}, and we expect that these results can be 
adapted to the conifold setup. In the hypermultiplet sector the problems
encountered in both the 4d and 5d case are the same: the 
hypermultiplet manifold is constrained to be quaternion-K\"ahler by
supersymmetry, but not much is known about such manifolds beyond
the 
case of a single hypermultiplet. 
Moreover, the work of \cite{OV}, who analyzed the rigidly supersymmetric
limit using symmetry arguments, shows that on the Higgs branch the 
conifold singularity is resolved by non-perturbative effects which are not
directly related to integrating out states which become massless at the
transition point.\footnote{In fact this is necessary
for the consistent description of conifold transitions in terms of an LEEA, 
since the extra states on the Higgs branch are long vector multiplets,
which in do not give rise to threshold corrections when integrated out.}
Due to these complications,  it is not known how to compute the 
hypermultiplet metric from string or M-theory. In \cite{CFL} we therefore 
decided to use the
simplest family of quaternion-K\"ahler spaces 
%
\be
X(n_H) = \frac{U(n_H,2)}{U(n_H) \times U(2)} \, .
\label{Wolf}
\ee
The data which uniquely determine the 5d LEEA are  
$X(n_H)$, the metric of the vector multiplet manifold 
(which, in principle,
can be computed exactly\footnote{For technical convenience we made a 
simple choice in \cite{CFL}.}), and a so-called gauging, i.e.,
the transformation of the charged
hypermultiplets under gauge transformations. 
In the next section we will discuss how the gauging is determined in terms
of the homology classes of the ${\cal C}_i$. 
The extension to type-IIA compactifications on $X$ is non-trivial, 
because the vector multiplet sector becomes more complicated, though
the other data remain unchanged. As
a first step we will present the naive dimensional reduction of the
5d model \cite{Mohaupt:2004pr} to four dimensions in section 3.

\section{Hypermultiplets charges}

Once the hypermultiplet metric is chosen to be \eqref{Wolf}, 
all the possible gaugings are determined by picking suitable subsets
of the Killing vectors, since gauge transformations must act by
isometries on the charged scalars. As shown in \cite{FS3,CFL} this 
boils down to specifying the charges carried by the hypermultiplets, 
which in turn are determined by the homology classes of the 
curves ${\cal C}_i$,
around which the D2/M2-branes are wrapped. In this section we will
review and extend these results to general conifold transitions.

All the 4d/5d-gauge fields $A^I_\mu$ descend from  higher dimensional three-form
gauge fields and are in one-to-one correspondence with the independent
harmonic 2-forms $\omega_I$ on $X$. Since the electric sources for the
three-form are D2/M2-branes, one obtains pointlike 4d/5d-sources 
by wrapping the 2-branes on holomorphic curves ${\cal C}$. The zero 
modes of such wrapped branes organize themselves into hypermultiplets \cite{Witten1}
whose electric charges $q_I$, $[{\cal C}] = q_I C^I$,  
are just the expansion coefficients of the homology class $[{\cal C}]$
of ${\cal C}$
with respect to a basis $C^I$ of $H_2(X,\mathbbm{Z})$
dual to $\omega_I$, $\int_{C^I} \omega_J = \delta^I_J$.
The charges corresponding
to the $N$ extra light hypermultiplets live in a sublattice of
$H_2(X,\mathbbm{Z})$, which is generated by the homology classes
$[{\cal C}_i]$ of the contracted curves ${\cal C}_i$. 
While the $[{\cal C}_i]$ are {\em effective} classes,
it  turns out to be convenient to use generators ${\rm C}^i = \pm [{\cal C}_i]$
for the sublattice, where not necessarily all of the signs are `$+$' 
(see below).\footnote{Note that only the effective homology classes 
correspond to submanifolds of $X$, while in general a homology class
 is an equivalence class of formal integer linear combinations
of submanifolds. In particular,  
if a class $C$ is effective, then  $-C$ is not. 
}

Let us now assume that by varying the vector multiplet 
scalars, which encode the K\"ahler moduli of $X$, we have reached
a point in moduli space where $N$ holomorphic curves ${\cal C}_i$ 
contract to zero volume.
Depending on the homology classes $[{\cal C}_i]$, this point may allow for a flop or conifold transition. 
If all ${\cal C}_i$ are in the same homology class $C=[{\cal C}_i]$,
we can perform a flop transition, which means that, after shrinking 
to zero volume, the ${\cal C}_i$ re-expand to finite volume, but now
belong to the homology class $\tilde{C}=-C$. The resulting new smooth
CY${}_3$ $\tilde{X}$ is topologically different from $X$. In particular
its triple intersection numbers $\tilde{C}_{IJK}$ which control the
vector multiplet couplings are different from those of $X$.
%
In the parametrization chosen in \cite{FS3} all the hypermultiplets
carry charge $+1$ with respect to the gauge field 
$A_\mu := q_I A^I_\mu$.\footnote{
Hypermultiplets are CP self-conjugate and therefore 
contain for each particle state also the anti-particle state of opposite
charge. In \cite{FS3} we parametrized the hypermultiplet scalars in 
terms of two complex scalar fields of opposite
charge, denoted $u$ and $v$,  and we defined the charge of the $u$-field 
to be `the charge' of the hypermultiplet.} 
We then found that the scalar potential gives the correct vacuum structure and mass matrix expected
from the underlying microscopic physics: an unlifted Coulomb 
but no Higgs branch. 

To have a conifold transition 
one needs $N$ collapsing cycles ${\rm C}^{i}$, which are subject
to $r$ linear homology relations, where $N>r>0$. The transition
works by contracting the $N$ holomorphic curves and re-expanding them
as $N$ special lagrangian submanifolds, which have 
real dimension 3. This transition  
relates a CY${}_3$ with Hodge numbers $h^{1,1}, h^{1,2}$ and
Euler number $\chi$ to a new smooth CY${}_3$ $\tilde{X}$ with
Hodge numbers $\tilde{h}^{1,1} = h^{1,1} - (N-r), 
\tilde{h}^{1,2} +r $ and Euler number $\tilde{\chi} = \chi - 2N$.

To explain the charge assignment, let us first discuss 
the case where $N$ is arbitrary and $r=1$ \cite{CFL}. The fact that there is
just one relation among the ${\cal C}_i$ implies that they
form the boundary of a single three-chain.
The resulting homology relation is
\be
{\rm C}^{1} + {\rm C}^{2} + \ldots + {\rm C}^{N} = 0 \, . 
\label{Relr1}
\ee
This can be used to solve for one of the cycles in terms of the others.
Without loss of generality we take the classes 
${\rm C}^{\alpha}$, $\alpha = 1, \ldots ,
N-1$
to be independent and solve for ${\rm C}^{N}$. Moreover, if we take 
the ${\rm C}^{\alpha}$ to be effective classes, 
${\rm C}^{\alpha} = [{\cal C}_\alpha]$,
then ${\rm C}^{N}$ is clearly not effective. We could work in terms of the 
effective class $-{\rm C}^N$, but we find it more convenient to have 
all the signs in (\ref{Relr1}) to be identical.
As a consequence of (\ref{Relr1}) the extra hypermultiplets are charged under 
$N-1$ (rather
than $N$) independent abelian gauge fields. Taking the independent
gauge fields to be the $A^\alpha_\mu$ corresponding to the ${\rm C}^{\alpha}$,
the charge of the $i$-th hypermultiplet under $A^\alpha_\mu$ is
$q_{i \alpha} = + \delta_{i\alpha}$ for $i=1,\ldots, N-1$ and
$q_{N \alpha} = -1$ for all $\alpha$. 
%

We now extend this result to the case where we have $r=N-1$ relations, with $N \ge 2$ arbitrary.
Here the $N$ classes $[{\cal C}_i]$ are the boundaries
of $N-1$ independent three-chains. Since each such three-chain must have
more than one boundary component, the homology relations take
the form
\be\label{RelrN-1}
{\rm C}^{1} + {\rm C}^{2} = 0 \;, \;\;\;
{\rm C}^{2} + {\rm C}^{3} = 0 \;, \;\;\;
\ldots \;,
{\rm C}^{N-1} + {\rm C}^{N} = 0 \;.
\ee
Thus all the cycles fall into just two different homology classes,
which differ by an overall sign. Taking, say, 
${\rm C}^{1}= {\rm C}^{3} = \ldots$, to be the effective class, 
the other class is not effective:
${\rm C}^{1}= {\rm C}^{3} = \ldots = - {\rm C}^{2} = - {\rm C}^{4} = \ldots$.
There is only one independent gauge field $A_\mu$, under which
the multiplets with $i=1,3,\ldots$ carry charge $+1$, while those with
$i=2,4,\ldots$ carry charge $-1$. By a straightforward generalization of
the results of \cite{CFL} one can show that this charge assignment
defines a unique gauging of a LEEA based on the hypermultiplet 
manifold (\ref{Wolf}). Note that the particular example of a conifold
transition with $N=4,r=3$ discussed in \cite{Huebsch} and in 
Appendix C of \cite{CFL} falls into this class.


Let us remark that in the intermediate cases where $N-1 > r > 1$ the homology relations analogous to \eqref{Relr1} and \eqref{RelrN-1} are not uniquely fixed by the values of $N$ and $r$ but depend on the underlying geometry. Thus they have to be found case by case for each transition. However, once these relations are known the rule outlined above provides a prescription how to translate these homology relations into a charge assignment which uniquely determines the gauging of the LEEA.

\section{Conifold transitions in four-dimensional cosmologies}
After discussing the construction of the general extended LEEA for flop and conifold transitions, we now turn to the particular conifold model investigated in \cite{Mohaupt:2004pr}. This model has the virtue that it has the minimal field contend necessary for describing a conifold. While it is not clear that this particular model has an explicit realization in terms of a string compactification, we nevertheless expect that it captures all the essential features of realistic conifold transitions occurring in a full-fledged M-theory compactification.

Our starting point is the 5d low energy effective action
\be\label{3.1}
S_{\rm cone}   =  \int \rmd^5 x \sqrt{-g} \left( - \frac{1}{2} R  
- \frac{1}{2} g_{XY} \partial_{\mu} q^X \partial^{\mu} q^Y 
- \frac{1}{2} g_{xy} \partial_{\mu} \phi^x \partial^{\mu} \phi^y  
- \pV(\phi, q) \, \right) , 
\ee
where the scalar field metrics and scalar potential are given by
\be\label{7.8}
g_{xy}  =   \frac{3 \left( 2 + \phi^2 \right)}{(2 - 3 \phi^2)^2 } \; , \quad 
g_{XY}  =  \frac{2}{\left( 1 - 2 q^2 \right)^2} \; , 
\quad
\mbox{and} \quad
%
\pV(\phi, q) =  \frac{\left( 48 \pi \right)^{2/3} \, \phi^2 \, q^2}{\left( 1 - 2 q^2 \right) \, \left(1 - \frac{3}{2} \phi^2 \right)^{2/3}} \; ,
\ee
respectively. This Lagrangian provides a consistent truncation of the low energy supergravity action describing a minimal conifold transition.
In order to obtain the corresponding model for the type-IIA string we perform a naive dimensional reduction on a circle. 
Neglecting the vector fields and dilaton arising from the reduction of the 5d space-time metric, we arrive at a 
4d model of a conifold transition.\footnote{This naive reduction also neglects the perturbative and non-perturbative corrections to the prepotential determining the vector multiplet sector of the resulting 4d effective action. We expect that these can be treated along the lines \cite{LMZ}.} 

Following \cite{Mohaupt:2004pr} we now illustrate the typical behavior of cosmological solutions in this framework. In this course we take the 4d space-time metric to be the flat ($k = 0$) Friedmann-Robertson Walker metric
%
$
\rmd s^2 = - \rmd t^2 + \e^{2 \alpha(t)} \, \left( \rmd r^2 + r^2 \rmd \Omega_2^2 \right)
$
%
with $a(t) := \e^{\alpha(t)}$ being the usual scale factor. Defining
$ T := \frac{1}{2} \, g_{XY} \, \qd^X \, \qd^Y + \frac{1}{2} \, g_{xy} \, \pd^x \pd^y \, $,
the gravitational and matter equations of motion take the form
\be \label{7.3a}
3 \ad^2  =  T +  \pV \, , \qquad \ddot{\alpha} = - T \, , 
\ee
and
\be \label{7.4}
\ddot{\phi}^x + \gamma^x_{~yz} \, \pd^y \, \pd^z +  3 \, \ad \, \pd^x +  g^{xy} \frac{\partial \pV}{\partial \phi^y }  =  
0 \, , \quad 
\ddot{q}^X + \Gamma^X_{~YZ} \, \qd^Y \, \qd^Z +  3 \, \ad \, \qd^X +  g^{XY} \frac{\partial \pV}{\partial q^Y }  =  0 \, , 
\ee
where $\gamma^x_{~yz}$ and $\Gamma^X_{~YZ}$ denote the Christoffel symbols of the metrics $g_{xy}$ and $g_{XY}$, respectively. In order to illustrate the behavior of the cosmological solution, we also introduce the Hubble parameter $H(t):= \frac{\dot{a}}{a} = \ad$ and the acceleration parameter $\ac(t) := - \, \frac{\ddot{a} \, a}{\dot{a}^2} = - \, \frac{\ddot{\alpha} + \ad^2}{\ad^2}$.  
Accelerated expansion of the space-time corresponds to $\ddot{a} > 0 
\Leftrightarrow \ac(t) < 0$.

\begin{figure}[p!]

\epsfxsize=0.4\textwidth
\begin{center}
\leavevmode
\epsffile{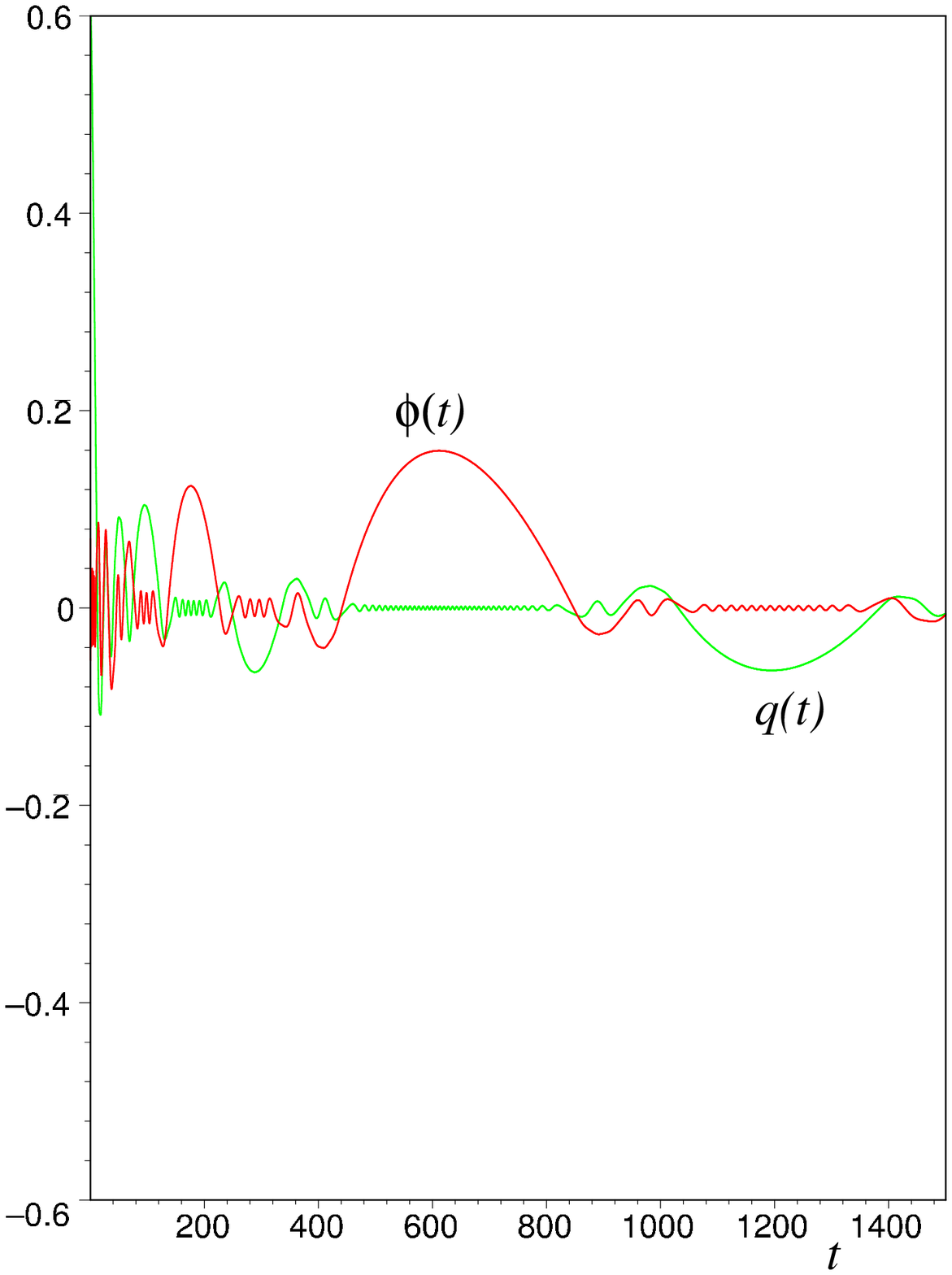} \,
\epsfxsize=0.4\textwidth
\epsffile{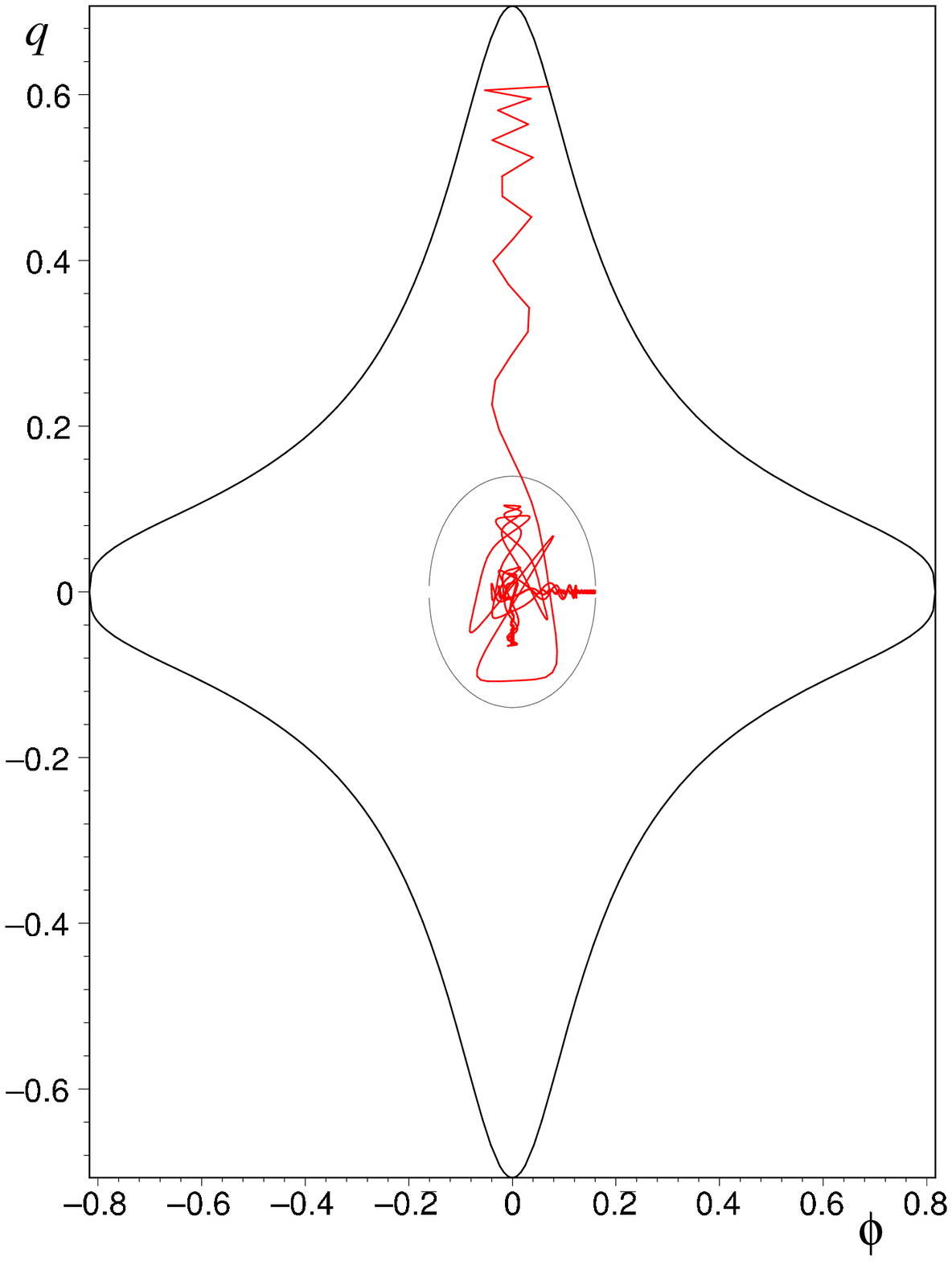}
\end{center}
\epsfxsize=0.4\textwidth
\begin{center}
\leavevmode
\epsffile{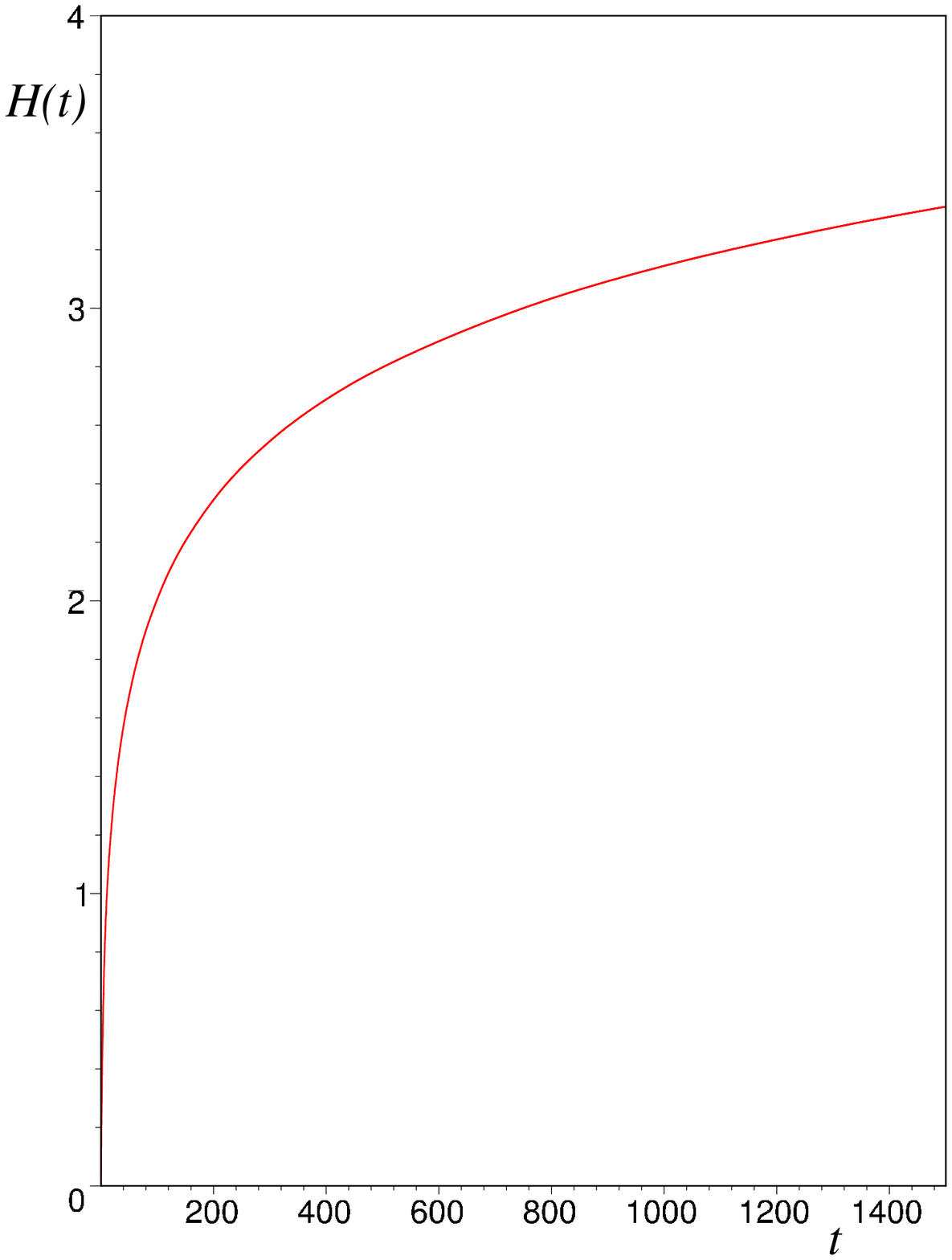} \,
\epsfxsize=0.4\textwidth
\epsffile{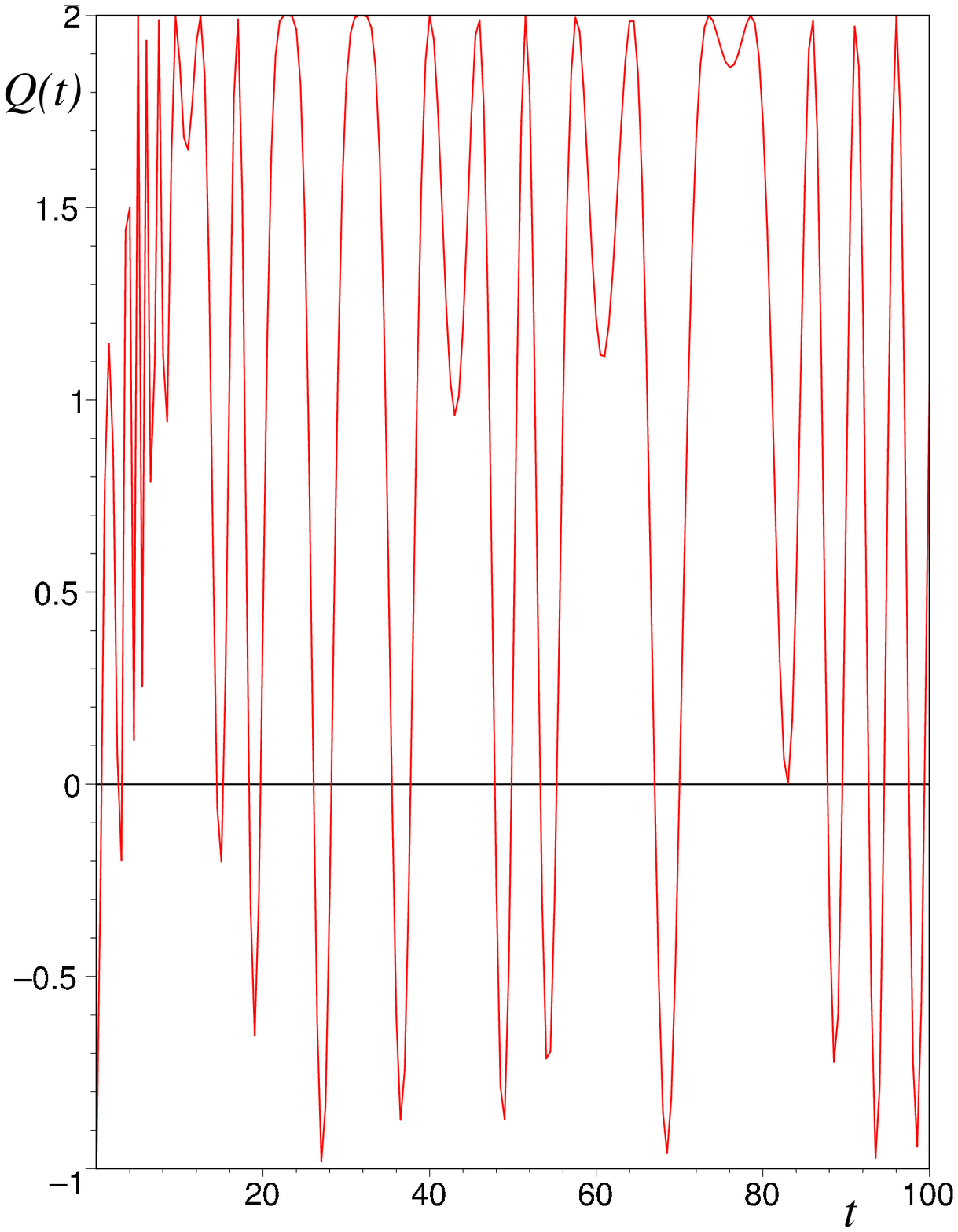}
\end{center}
\parbox[c]{\textwidth}{\caption{\label{eins}{\footnotesize Typical behavior of a solution of the equations of motion \eqref{7.3a} and \eqref{7.4} within an expanding universe. The upper left diagram shows the evolution of $\phi(t), q(t)$ with respect to cosmological time. For $0 \lesssim t \lesssim 100$ and $1050 \lesssim t \lesssim 1150$ the trajectory runs along the Higgs branch while for $550 \lesssim t \lesssim 700$ it evolves along the Coulomb branch, indicating that the solution undergoes several conifold transitions. The upper right diagram displays the same trajectory projected to the $\phi$-$q$-plane. Here the bold black line illustrates the initial energy of the scalar fields while the gray circle illustrates our choice of the central region around the conifold point $\phi=q=0$. After the initial approach the trajectory gets trapped in this region. The lower left and lower right diagrams show the Hubble parameter $H(t)$ and the acceleration parameter $Q(t)$, respectively.}}}
\end{figure}

We will now discuss a typical solution of the equations of motion \eqref{7.3a}, \eqref{7.4} corresponding to an expanding universe. The evolution of the scalar fields $\phi(t)$ and $q(t)$ is shown in the upper left and upper right diagram of Fig.\ \ref{eins}, while $H(t)$ and $Q(t)$ are shown in the lower left and lower right diagram, respectively. 
%

Looking at the time-dependence of $\phi(t), q(t)$ (upper left diagram) we observe that, analogous to the 5d case, conifold transitions are realized
dynamically.
Indeed, taking the order parameter for the transition to be $\xi(t) := |q(t)|/|\phi(t)|$ and adopting the criterion $\xi(t) \ge 100$ ($\xi(t) \le 0.01$) for the solution to evolve along the Higgs (Coulomb) branch, the solution shows multiple transitions between these branches.  
Projecting the solution to the $\phi$-$q$-plane in the upper right diagram 
we see that it 
gets trapped in the vicinity of the conifold point at $\phi = q = 0$. Analogous to the 5d case, this trapping is caused by an interplay between Hubble friction and the shape of the scalar potential which always pushes the solution back to the central region. In the 4d case the Hubble friction is slightly less effective due to the decreased numerical factor in front of the Hubble friction terms $\ad \qd$ and $\ad \pd$. 

The lower left diagram shows that $H(t)$ increases monotonically in
time. After a period of rapid increase ($0 \le t \le 300$, say) its numerical
value approaches a plateau value for late times. The numerical value of this
plateau depends on the particular initial conditions chosen but typically
gives $H_{\rm Plateau} \approx 3$. The acceleration parameter (lower right
diagram) oscillates rapidly between $-1 \le \ac(t) \le 2$. The diagram thereby
shows a number of short periods of accelerated expansion (corresponding to
$\ac(t) < 0$) which, however, are not pronounced enough to give rise to a
significant expansion of the space-time. Our results are consistent
with what one expects from existing work on 
cosmological solutions in the vicinity of flop or conifold transitions 
\cite{flop,FS4,Mohaupt:2004pr,Lukas:2004du}.


\end{document}